\documentclass{article}
\usepackage[utf8]{inputenc}
\usepackage{authblk}
\usepackage{breqn}
\usepackage{amsfonts}
\usepackage{amsmath}
\usepackage{appendix}
\title{The periodically driven electron in a quantum well with two characteristic curvatures - redux}
\author{Rafael Bautista-Mena }
\affil{Universidad de los Andes}

\begin{document}

\maketitle

\begin{abstract}
I develop the solution to the problem of an electron confined in a composite quadratic well subject to a simple, external periodic force. The method of solution illustrates several of the basic techniques useful in formally solving the one-dimensional, time-dependent Schrodinger equation. One of the aims of this exercise is to see how far is it possible to push analytics, before plowing into numerical methods. I hope this presentation of the problem may result useful to others seeking to gain additional experience in the details of solving the time-dependent Schrodinger equation in one space dimension.

Keywords: Composite quantum wells, external field, formal solution 

\end{abstract}

\section{Introduction}
\label{sec:intro}
What is now decades ago, an acquaintance \cite{NeilF93} suggested to me the following problem: given a one-dimensional,  two-piece harmonic well, driven by a single-frequency external wave, what sort of resonant behaviour could be expected to arise? Since I haven't kept up to date with the literature on this subject, I can only assume that plenty has happened since then, and that the attempt that follows may still be of interest mainly in the sense that adds yet another approach to the problem.

The eigenvalue equation for the ground Hamiltonian is
\begin{equation} \label{Eq1}
    H_{0}\psi_{k}=\epsilon_{k}\psi_{k}
\end{equation}
The confining potential is a composite quadratic well of the form
\begin{equation}
    V(x)=\begin{cases}
    \frac{1}{2} k_{1}x^{2} & \text{$x\geq 0$}\\
   \frac{1}{2} k_{2}x^{2} & \text{$x\leq 0$}
    \end{cases}
\end{equation}
In this expression, $k_{1},k_{2}>0$. The problem of finding the energy levels for $H_{0}$ has been solved \cite{Glasser79}.

The ground-plus-interaction Hamiltonian is
\begin{equation}\label{H1}
    H_{1}=H_{0}+\gamma x \cos{\omega t}
\end{equation}
In this expression $\gamma$ represents the amplitude of the external periodic electric field times the particle's charge.

Let's denote $\Phi\left(x,t\right)$ the wave function that solves the corresponding time-dependent Schrodinger equation:
\begin{equation} \label{Eq2}
    H_{1}\Phi\left(x,t\right)=i\hbar\frac{\partial\Phi\left(x,t\right)}{\partial t}
\end{equation}
With the boundary conditions $\Phi\left(\pm \infty,t\right)=0$.

The initial condition is somewhat more delicate, due to considerations about the moment of ``switching-on'' of the forcing field. As a tentative choice, we will set at some arbitrary $t=t_{0}$, a moment well after the initial onset of the forcing electric field.

The proposed form of the solution is
\begin{equation} \label{phi0}
    \Phi\left(x,t\right)=\sum_{k}{\varphi_{k}(t)\psi_{k}(x)\exp{(-i\omega_{k} t)}}
\end{equation}
In this expression $\omega_{k}\equiv \epsilon_{k}/\hbar$. Since the $\psi_{k}$ constitute an orthonormal set, and comply with the boundary conditions, this form of the solution allows for focusing on the time-dependent part.

Substituting (\ref{phi0}) into (\ref{Eq2}) produces, after using (\ref{Eq1}), some simplifications:
\begin{equation}
\begin{split}
\gamma x \cos{\omega t}\sum_{k}{\varphi_{k} (t)\psi_{k}(x)\exp{(-i\omega_{k} t)}} \\ =i\hbar\sum_{k}{\frac{\partial\varphi_{k} (t)}{\partial t}\psi_{k}(x)\exp{(-i\omega_{k} t)}}
\end{split}
\end{equation}
Taking the $m$-th component of this equation, leaves
\begin{equation}
\begin{split}
    \gamma\cos{\omega t}\sum_{k}{x_{mk}\varphi_{k}(t)\exp{(-i\omega_{k} t)}} \\
    = i\hbar\frac{\partial\varphi_{m} (t)}{\partial t}\exp{(-i\omega_{m} t)}
    \end{split}
\end{equation}
Let $\Tilde{X}$ be the dipole matrix whose elements are $x_{mk}$. Let $\hat{\varphi}(t)$ be a vector whose elements are $\varphi_{k}\exp{(-i\omega_{k} t)}$. Let $\Tilde{\omega}$ be a diagonal matrix whose elements are the $\omega_{k}$. Then, the last equation may be rewritten as
\begin{equation} \label{Eq}
\gamma\cos{\omega t} \Tilde{X}\hat{\varphi}=i\hbar\frac{d\hat{\varphi}}{dt}-\hbar\Tilde{\omega}\hat{\varphi}
\end{equation}
This equation may be expanded in terms of the basis vectors $\hat{v}_{k}$ and corresponding eigenvalues $\lambda_{k}$ associated to the dipole matrix $\Tilde{X}$.
\begin{equation} \label{eigenX}
    \Tilde{X}\hat{v}_{k}=\ell\lambda_{k}\hat{v}_{k}
\end{equation}
The parameter $\ell$ is positive and has dimension of length. It is constructed out of the dynamical parameters in the ground Hamiltonian. Denoting $m$ the mass of the particle, it is given by
\begin{equation}
    \ell^{2}=\frac{\hbar}{\sqrt{m\kappa}}
\end{equation}
where $\kappa\equiv\sqrt{k_{1}k_{2}}$.

From now on, the components $x_{lm}$ in the dipole matrix are re-scaled, so that they are measured in units of $\ell$.
Equation (\ref{Eq}) may be rewritten now in terms of its components in the dipole basis, starting from the expansion
\begin{equation} \label{Eq3}
    \hat{\varphi}(t)=\sum_{k}{f_{k}(t)\hat{v}_{k}}
\end{equation}
Substituting (\ref{eigenX}) and (\ref{Eq3}) into (\ref{Eq}), and using matrix notation, (\ref{Eq}) acquires the form
\begin{equation}\label{Eq4}
    \gamma\cos{\omega t}\Tilde{\Lambda}\hat{f} =i\hbar\frac{d\hat{f}}{dt}-\hbar\Tilde{\Omega}\hat{f}
\end{equation}

Where $\Tilde{\Lambda}$ is a diagonal matrix formed with the dipole eigenvalues $\lambda_{k}$ and $\Tilde{\Omega}$ is the dipole basis representation of the diagonal matrix $\Tilde{\omega}$, built from the $\omega_{k}$, which no longer looks diagonal.

One last transformation to a new dependent variable leads to the final form of the problem:
\begin{equation}
\hat{f}(t)=\exp{(-i\beta\sin{\omega t}\Tilde{\Lambda})}\hat{q}(t)
\end{equation}
where $\beta\equiv\gamma\ell/\hbar\omega>0$ is a dimensionless quantity. With this, equation (\ref{Eq4}) takes the standard form
\begin{equation} \label{Eq5}
i\frac{d\hat{q(\xi)}}{d\xi}=\Tilde{K}(\xi)\hat{q}(\xi)
\end{equation}
In the expression above, $\xi \equiv \omega t$.
The time evolution matrix is given by
\begin{equation} \label{K}
    \Tilde{K}(\xi)=\frac{1}{\omega}\exp{(i\beta\sin{\xi}\Tilde{\Lambda})}\Tilde{\Omega}\exp{(-i\beta\sin{\xi}\Tilde{\Lambda})}
\end{equation}
The formal solution of equation (\ref{Eq5}) is then
\begin{equation} \label{solution}
\hat{q}(\xi)=\exp{\left(-i\int_{\xi_{0}}^{\xi}{\Tilde{K}(z)dz}\right)}\hat{q}(\xi_{0})
\end{equation}


\section{TIME EVOLUTION}
Starting from (\ref{solution}), it is possible to make some progress in reducing it to more recognizable terms through a third transformation, to the set of eigenvectors of the integral of the time evolution matrix:
\begin{equation}\label{U}
    \Tilde{U}(\xi,\xi_{0})=\int_{\xi_{0}}^{\xi}{\Tilde{K}(z)dz}
\end{equation}
With this integration, (\ref{solution}) adopts the form
\begin{equation}\label{Uevol}
    \hat{q}(\xi)=\exp{\left[-i\Tilde{U}(\xi,\xi_{0})\right]}\hat{q}(\xi_{0})
\end{equation}

The practical problems posed by obtaining an operational approximation for equation (\ref{Uevol}) can be approached by considering the eigenvalue problem
\begin{equation}\label{eigenU}
    \Tilde{U}(\xi,\xi_{0})\hat{u}_{l}(\xi)=\mu_{l}(\xi)\hat{u}_{l}(\xi)
\end{equation}
From (\ref{K}) and (\ref{U}) is easy to see that $\Tilde{U}=\Tilde{U}^{\dagger}$ . Therefore, the eigenvalues  are real. Once $\Tilde{U}(\xi,\xi_{0})$ is known, this problem can be successfully tackled for a chosen table of values $\xi>\xi_{0}$. This scheme, however, runs into some practical difficulties. Notice, for instance, that, from (\ref{U}), $\Tilde{U}(\xi_{0},\xi_{0})=\Tilde{0}$, therefore, it is necessary to define through some suitable limiting process the corresponding set $\hat{u}_{l}(\xi_{0})$. Some of the steps necessary toward achieving a set of tables for the $u_{l}(\xi)$, and the corresponding $\mu_{l}(\xi)$, from (\ref{eigenU}) are discussed in the Appendix.

Notice that the solution of (\ref{eigenU}) implies finding a complete, normalized set $\hat{u}_{l}(\xi)$ for each tabulated value of $\xi$. Therefore, for each $\xi$, we have the following identity
\[
\sum_{l=1}^{\infty}{\hat{u}_{l}(\xi) \hat{u}_{l}^{\dagger}(\xi)}=\Tilde{1}
\]
We may insert this identity in (\ref{Uevol}) and then use (\ref{eigenU}) to arrive at
\begin{equation}\label{formalSol}
    \hat{q}(\xi)=\sum_{l=1}^{\infty}{\hat{u}_{l}^{\dagger}(\xi) \hat{q}(\xi_{0})\exp{\left[-i\mu_{l}(\xi)\right]}\hat{u}_{l}(\xi)}
\end{equation}

The initial state $\hat{q}(\xi_{0})$ is necessarily the result of some preparation, which may be represented using the dipole basis,
\begin{equation}
    \hat{q}(\xi_{0})=\sum_{l=1}^{\infty}{q_{l}^{0}\hat{v}_{l}}
\end{equation}

With this prescription, the problem of computing the time evolution of the wave function is formally solved. Walking all the steps back from (\ref{formalSol}) up to (\ref{phi0}) recovers the definitive shape of the proposed solution:
\begin{equation} \label{Phi}
    \Phi(x,t)=\sum_{k=1}^{\infty}\sum_{m=1}^{\infty}{\exp{\left[-i\beta \lambda_{k} \sin{\xi} \right]q_{k}(\xi)\psi_{m}(x)v_{k,m}}}
\end{equation}
where $v_{k,m}$ is the $m$ component of eigenvector $k$.

\section{CONCLUSION}
Of course, despite having arrived at (\ref{Phi}), it is still far from getting to actual numerical results for some particular set of parameter values of the Hamiltonian (\ref{H1}). Given that all sequences of eigenvalues and vector components are infinite, a compromise about, for instance, how many rows and columns of the dipole matrix $\Tilde{X}$ are going to be calculated has to be made. Enough rows and columns to get matrix elements that, hopefully, vanish quickly with increasing size of the matrix.
However, finding a particular solution for the wave function is not the main aim of this analysis. The key to answering the question posed at the start of this paper is completely contained in the shape of $\Tilde{U}(\xi,\xi_{0})$.

It is possible to write $\Tilde{U}(\xi,\xi_{0})$ in a closed, compact form:
\begin{equation}\label{Key}
    \Tilde{U}(\xi,\xi_{0})=\sum_{l=0}^{\infty}\sum_{m=0}^{\infty}{\Omega_{lm} I(\xi,\xi_{0}|\alpha_{lm})\hat{v}_{l}\hat{v}_{m}^{\dagger}}
\end{equation}
Where  $\alpha_{lm} \equiv \beta (\lambda_{l}-\lambda_{m})$. $I(\xi,\xi_{0}|\alpha_{lm})$ is a Fourier series over all the harmonics of $\omega$, and the coefficients in the series turn out to be proportional to the Bessel functions of the first type with argument $\alpha_{lm}$. The precise details for each of the terms in this equality are given in the Appendix. The important point to be made is that (\ref{Key}) is the equivalent of the solution to the classical, forced harmonic oscillator, where the behaviour of the coefficients of the sine and cosine functions, as $\omega$ varies, shows peaks, and other features, that point to the existence of resonances. The only convincing manner of showing what the resulting behaviour could be in the case of (\ref{Key}) is through an actual numerical example.


\pagebreak
\appendix{APPENDIX: LOOKING AT THE INTEGRAL}

The definitions for $\Tilde{K}$ and $\Tilde{U}(\xi,\xi_{0})$, when taken together and expressed explicitly in terms of the dipole eigenvalues, produce matrix elements of the form
\begin{equation}\label{Ulm}
    U_{lm}(\xi,\xi_{0})=\frac{1}{\omega}\int_{\xi_{0}}^{\xi}{\Omega_{lm}\exp{\left[i\beta\left(\lambda_{l}-\lambda_{m}\right)\sin{z}\right]}dz}
\end{equation}
The integrand in ($\ref{Ulm}$) has the general form
\begin{equation} \label{I}
    I(\xi,\xi_{0}|\alpha)=\int_{\xi_{0}}^{\xi}{\exp{\left(i\alpha\sin{z}\right)}dz}
\end{equation}
with $\alpha$ a real-valued quantity.

The expansion of the exponential in powers of $\alpha$ produces a series which, in compact form, may be written by first defining the single-argument version
\begin{equation}\label{Iexpand}
   I(\xi|\alpha) =J_{0}(\alpha)\xi+\sum_{m=0}^{\infty}{\left[\left(\frac{\alpha}{2}\right)^{2(m+1)}\eta_{m}(\xi)
    -i\left(\frac{\alpha}{2}\right)^{2m+1}\phi_{m}(\xi)\right]}
\end{equation}
where the $\eta_{m}$ and $\phi_{m}$ are given by
\begin{equation}\label{eta}
    \eta_{m}(\xi)=\frac{1}{(2m+2)!}\sum_{s=0}^{m}{(-)^{s}\binom{2(m+1)}{s}\frac{\sin[2(m+1-s)\xi]}{2(m+1-s)}}
\end{equation}
And
\begin{equation}\label{phi}
    \phi_{m}(\xi)=\frac{1}{(2m+1)!}\sum_{s=0}^{m}{(-)^{s}\binom{2m+1}{s}\frac{\cos[2(m-s)+1)\xi]}{2(m-s)+1}}
\end{equation}

These expressions, besides showing the explicit dependence of matrix $\Tilde{U}$ on $\xi$, are useful to write $I(\xi,\xi_{0}|\alpha)$ in a way that allows for the evaluation of convergence properties in terms of powers of the argument $\alpha$. As it happens, (\ref{I}) turns out to be the integral of an exact differential
\begin{equation}
    I(\xi,\xi_{0}|\alpha)=I(\xi|\alpha)-I(\xi_{0}|\alpha)
\end{equation}

Expressions for (\ref{eta}) and (\ref{phi}) tend to obscure the explicit contribution of each harmonic to the sum. Some control on the computational process is gained with the rewriting of (\ref{Iexpand}) as a Fourier series.

For (\ref{phi}), the associated orthogonal functions are of the form $\cos{[(2m+1)\xi]}$ and for (\ref{eta}) the associated functions are $\sin{(2m\xi)}$. $m$ is any non-negative integer. The Fourier coefficients turn out to be proportional to the cylindrical Bessel functions of the first kind, leading to the following equivalences
\begin{equation}\label{Even}
    \sum_{m=0}^{\infty}{\left(\frac{\alpha}{2}\right)^{2(m+1)}\eta_{m}(\xi)}=\sum_{k=0}^{\infty}{\frac{1}{2(k+1)}J_{2(k+1)}(\alpha)}\sin{[2(k+1)\xi]}
\end{equation}
As a shorthand, the right-hand series is $E(\xi|\alpha)$.
And
\begin{equation}\label{Odd}
    \sum_{m=0}^{\infty}{\left(\frac{\alpha}{2}\right)^{2m+1}\phi_{m}(\xi)}=\sum_{k=0}^{\infty}{\frac{1}{2k+1}J_{2k+1}(\alpha)}\cos{[(2k+1)\xi]}
\end{equation}
As a shorthand, the right-hand series is $P(\xi|\alpha)$.

With these results and conventions, the Fourier representation of $I(\xi|\alpha)$ is
\begin{equation}\label{IFourier}
    I(\xi|\alpha)=J_{0}(\alpha)\xi+E(\xi|\alpha)-i P(\xi|\alpha)
\end{equation}
The result that the integrals defining the matrix $\Tilde{U}(\xi,\xi_{0})$ turn out to come from exact differentials allows the writing of a closed expression. Setting $\alpha=\beta (\lambda_{l}-\lambda_{m})$ in (\ref{IFourier}), first define the matrix
\begin{equation}
    \Tilde{U}(\xi)=\sum_{l=0}^{\infty}\sum_{m=0}^{\infty}{\Omega_{lm} I(\xi|\alpha_{lm})\hat{v}_{l}\hat{v}_{m}^{\dagger}}
\end{equation}
With this, (\ref{U}) becomes
\begin{equation}
    \Tilde{U}(\xi,\xi_{0})=\Tilde{U}(\xi)-\Tilde{U}(\xi_{0})
\end{equation}

In this way, the solution to the eigenvalue problem posed in equation (\ref{eigenU}) becomes a matter of procedures in numerical analysis.

The Fourier representation (\ref{IFourier}) has one advantage: in the sums (\ref{Even}) and (\ref{Odd}), $k$ tells both the order of harmonics in $\omega$ and the lowest order of power in $\alpha$ for any term. When it comes to doing numerical approximations, a decision to go up to, say, order $\alpha^{5}$, implies that effects due to harmonics also reach up to harmonic $5\omega$.


\begin{thebibliography}{}
\bibitem{NeilF93}
\textit{N. F. Johnson and J. E. Vargas; Optical absorption in quantum wells with two characteristic curvatures. Appl. Phys. Lett. 62, 627 (1993).}
\bibitem{Glasser79}
\textit{M. L. Glasser, Determining the energy levels of composite potential wells. Am. J. Phys. 47, 738 (1979).}
\end{thebibliography}
\end{document}